\documentstyle[aps,prb,epsfig]{revtex}
\begin{document}
\twocolumn[\hsize\textwidth\columnwidth\hsize\csname
@twocolumnfalse\endcsname
\title{Magnetoresistances  observed  by  decomposition  of  the  magnetic   moment in
La$_{1-x}$Ca$_x$MnO$_3$ films$^{\ast}$}
\author{Hyun-Tak Kim$^{\ast\ast}$ and Kwang-Yong Kang}
\address{Telecom. Basic Research Lab., ETRI, Taejon 305-350, Korea}
\author{Eun-Hee Lee}
\address{Nuclear Material Technology Developments, KAERI, Taejon 305-600, Korea}
\maketitle{}
\begin{abstract}
A ferromagnetic phase, characterized by electron carriers and a high
temperature colossal magnetoresistance
 (HTCMR) dependent on the magnetic moment, and a semiconducting phase,
 characterized by hole carriers and a low temperature CMR (LTCMR),
are observed in La$_{1-x}$Ca$_{x}$MnO$_3$ thin  films by the  van
der Pauw  method. The LTCMR is much more sensitive to the
magnetic field than the HTCMR. In the ferromagnetic phase for
films with anisotropic moments in two dimensions, a remnant
resistivity of the order of 10$^{-8} ~{\Omega}m$   is observed up
to   100 K   and increases exponentially with both a temperature
up to $T_c$ and a magnetic field above one  Tesla (a positive
magnetoresistivity). We found that the ferromagnetic phase below
$T_c$ is in a polaronic state with a polaronic
 mobile conduction, and the carrier density dips near $T_c$. For
resistances measured
 by the four-probe method with line electrodes, low temperature information of the HTCMR
is not revealed. The van der Pauw method is more effective for
the resistance measurement of a magnetic material than the
four-probe method.
\\
\\
\end{abstract}
]
\section{INTRODUCTION}
Since the discovery of colossal  (or giant) magnetoresistances
(C  (or G) MR)$^{1-3}$ for hole-doped manganese oxides
La$_{1-x}$(Ca or  Sr)$_{x}$MnO$_3$, (LCMO or  LSMO), the
correlation between conductivity and
 ferromagnetism and  the  conduction mechanism   in the  ferromagnetic state   have been
intensively studied.
The correlation   is related  to the   strong Hund  coupling between   $t_{2g}$ spins  and
${\epsilon}_g$ electrons.
 For over 40  years, the coupling between  the charge  and the  spin has been  considered
 very  important.$^{4,5,6}$  However, some   researchers have  indicated that   the coupling
between  the  charge   and the   spin  is   not strong   enough
to  describe   the CMR correctly.$^{7,8}$ They proposed local
lattice distortions near $T_c^{7,8}$  and large oxygen isotope
effects on $T_c$.$^{9}$ Furthermore, the CMR mechanism is not yet
fully understood.$^{10}$

With regard to the  conduction mechanism below $T_c$  for LCMO or
LSMO, de Gennes proposed a second  double-exchange model  which
assumes  that the  charge carriers  are itinerant.$^{6}$ However,
a polaronic state was  suggested from observations of the
resistivity and the thermoelectric power for LCMO at $x$=0.12 and
0.15, showing semiconducting behavior below $T_c$.$^{11}$  Two
phases, one metallic and  the other semiconducting,  were
separated by observing magnetoresistances for  LCMO thin
films.$^{12}$ A detailed review on phase separation in CMR
materials was given.$^{13}$ However, it was also suggested that
the LSMO is a half metal with both metallic and semiconducting
electronic structures below $T_c$.$^{14}$ Thus, the conduction
mechanism and the electronic structure for LCMO and LSMO remain
unclear.

In this paper, we observe two distinct phases in LCMO films by the
van der Pauw method and measure the remnant resistivity, its
magnetic field dependence (a positive MR), and the Hall
coefficient in the ferromagnetic phase. The temperature dependence
of Hall coefficients is presented and the results are discussed
on the basis of other published results. Furthermore, for the
resistance measurement of magnetic materials, the four-probe
method with line electrodes is compared with the van der Pauw
method with point electrodes.

\section{Experiment}
Thin films  were deposited  with  a La$_{0.67}$Ca$_{0.33}$MnO$_3$
ceramic target  on (100)MgO substrates  at a  substrate
temperature of  700$^{\circ}$C in oxygen by laser ablation. The
deposited  LCMO films  showed semiconducting behavior  for
resistance before annealing. This may be attributed to the
lattice mismatch between the film and the MgO substrate. The
deposited LCMO films were annealed at 900$^{\circ}$C in oxygen in
a furnace       and        showed ferromagnetic behavior. The
La$_{0.70}$Ca$_{0.30}$Mn$_{0.97}$O$_3$ (LCMO2/12H),
La$_{0.63}$Ca$_{0.36}$Mn$_{1.17}$O$_3$ (LCMO3/1H), and
La$_{0.67}$Ca$_{0.33}$Mn$_{1.07}$O$_3$ (LCMO8/10H) films were
annealed for twelve, one, and ten  hours, respectively.  Their
dimensions were  3$\times$4 mm$^2$,  3.3$\times$3.7 mm$^2$, and
3$\times$3 mm$^2$, respectively, and the thicknesses were 3800
$\AA$, 2500 $\AA$, and 4080 $\AA$,  respectively. In our notation
LCMO$n$/$m$H, $n$ and  $m$ denote the number of the  samples and
annealing time,  respectively. Chemical contents of the films
were revealed by electron  probe microanalysis. Ca was distributed
homogeneously in the films. The films were oriented along the
c-axis by an X-ray diffractometer. After annealing films for more
than 4 hours, the full width at half-maximum  of the X-ray
diffraction peaks was less than that before annealing. The oxygen
content and crystalline order might be optimized by gradually
varying the annealing time. For measuring magnetoresistances,
electrodes were attached to  four edges of the film with a square
structure and denoted by marks  A, B, C and D in a clockwise
manner. When the current flows from A to B (x-axis), the voltage
is measured at D and C ($\rho_{xx}$);  when the current flows
from B to C (y-axis), the voltage is measured at A and D
($\rho_{yy}$). The  axis direction is arbitrary for any film. The
above procedure is
 known as the van der  Pauw
method. Hall coefficients were measured along  the diagonal
direction orthogonal to the current flow.$^{15}$ The magnetic
field was applied along the c-axis of the  films in  a SQUID  of
Quantum Design Co..  We used a Hewlett Packard nanovoltmeter in
this study.

\section{RESULTS and Discussion}
In a previous  paper,$^{12}$ we  found a  low-temperature colossal
magnetoresistance (or resistivity) (LTCMR) with peaks at 90 K and
a high-temperature colossal magnetoresistance (or  resistivity)
(HTCMR)  with  peaks  at   $T_c$=210 K in the LCMO3/1H film, as
shown  in Fig. 3 (a) and (b). Here, the HTCMR depends  on the
ferromagnetic moment. We found that with an increasing annealing
time the LTCMR decreases  and the  HTCMR develops; this is due to
the canting behavior   of the moment.$^{12}$  The  two CMR's  are
characteristic of the respective phases in assuming two phases.
The two phases were not distinguishable by X-ray diffraction
because of the high crystallinity of the thin films.

Figure 1 shows resistivities measured by the van  der Pauw method
with 50 ${\mu}$A at 0 and 5  Tesla for the LCMO8/10H  film. The
resistivities with  peaks at $T_c$=170 K and   200    K correspond
to    the   $x$-HTCMR   induced    by   the $x$-component
${\overrightarrow{M_x}}$  of  the projection
${\overrightarrow{M_{xy}}}$ on to the $x-y$ plane of a magnetic
moment ${\overrightarrow{M}}$. This is because the peak
temperatures agree closely with the temperature where
${\overrightarrow{M_z}}$ goes to zero, as shown in the inset of
Fig. 1. Below 100 K, remnant resistance voltage and resistivity
were measured at 0 Tesla and were found to be, on average,
3${\mu}V$ and 10$^{-3}~m{\Omega}cm$ (or 10$^{-8}~{\Omega}m$),
respectively. These values are regarded as constant with
temperature, although there is  low-level noise. The resistivity
is of the same  order as that  for a Y$_{0.7}$Sr$_{0.3}$MnO$_3$
thin film.$^{16}$ The resistivity below 100 K increases slightly
at 5 Tesla, as indicated by  A in the figure.  The logarithmic
resistivities  are  linear from 100  K to  $T_c$, which indicates
 nearest-neighbour hopping of polarons,  as suggested by Hundley  et al.$^{17}$. For the  resistivity
measured at 0 Tesla,  in assuming ${\rho}{\propto}exp(-E/k_BT)$
which is different  from the hopping term for semiconductors, $E$
is estimated at 157 $meV$ below $T_c$. This indicates that charges
below the Fermi  energy contribute to conduction. The mechanism
of the conduction is  still unclear below 100 K. The $y$-HTCMR
peaks D and E in  Fig. 1 are overlapped by the negative LTCMR
values on the order of 10$^{1}$ - 10$^{0}~m{\Omega}cm$ of peaks B
and C below 150 K. Therefore, the $y$-HTCMR cannot be determined
exactly and might not be equal  to the $x$-HTCMR. Peaks at B and
D are indistinguishable due to overlapping H and LTCMR. Generally,
the HTCMR decreases as the magnetic moment increases.$^{17,18}$
Here, the $y$-HTCMR is measured along the y-axis and induced by
${\overrightarrow{M_y}}$.

In  order  to  obtain  the  Hall   voltage of   the HTCMR,   we
made a LCMO2/12H  film. Magnetoresistance voltages ($V_{xx}$ and
$V_{yy}$) and the Hall voltage $V_{xy}$ were measured at 5  Tesla
from 5 to 400 K by the van der Pauw method, as shown in Fig.
2(a). The resistivities with a peak at 160 K correspond to the
HTCMR revealed by the magnetic moment, as shown elsewhere$^{12}$.
The LTCMR was undetectably low.  Two HTCMRs nearly agree, which
indicates that  the magnetic moment is nearly isotropic
(${\overrightarrow{M_x}}{\approx}{\overrightarrow{M_y}}$). Below
50  K, resistivities   are of  the order of  10$^{-1}$ -
10$^{-2}~m{\Omega}cm$, which  are much larger than the order of
10$^{-8}~m{\Omega}cm$ for LCMO8/10H in Fig. 1. These larger
resistivities occur because (1) the $x$-HTCMR when
${\overrightarrow{M_x}}={\overrightarrow{M_y}}$ is larger than the
$x$-HTCMR when ${\overrightarrow{M_x}}>{\overrightarrow{M_y}}$,
and  (2) the positive  MR competes with the negative LTCMR
depressed undetectably at low temperatures.

The Hall coefficient, $R_H$, and the carrier density measured at
5 Tesla for the LCMO2/12H film are  shown in the inset of Fig. 2
(b). The negative value of  the $R_H$ indicates that carriers  are
electrons which are regarded as   $d$-electrons. The $R_H$ shows
two peaks, and the   peak at 140   K corresponds to $T_c$ in the
resistivity in Fig. 2. The transition range from 140 to 230 K is
due to the Hall voltage variation near $T_c$, as shown in Fig. 2
(a). Above 230 K in the paramagnetic region, the $R_H$ decreases
with an increasing temperature. This is the same result obtained
by Jaime et al.$^{19}$, suggesting small polaron conduction.
Below $T_c$=140 K in the ferromagnetic region with a decreasing
temperature, the $R_H$ decreases, which indicates that the carrier
density increases as shown in Fig. 2 (b). The increase of the
carrier density results in a decrease of resistivity with a
decreasing temperature, which is not characteristic of the
metallic state but instead  is suggestive of a condensed state
with a polaronic conduction. This suggestion is consistent with
the results obtained by optical methods$^{20,21}$ and a
photoemission experiment$^{20}$. At $T_c$, the $R_H$  has a peak,
indicating a carrier density collapse which is the cause of the
HTCMR,  as suggested by Alexandrov and Bratkovsky$^{23}$. The
carrier density of 1.3${\times}$10$^{18}$~cm$^{-3}$  at 5 K was
calculated by assuming that the  effective mass is  equal to the
bare electron mass. The absolute value of  the carrier density is
uncertain because  the effective mass  of the carrier could not be
evaluated accurately. In  the polaron case the effective mass is
larger than  the  bare mass;  moreover,  for the   LCMO9/15H a
negative $R_H$ was also observed. The negative $R_H$ is
consistent with negative thermopower as measured in a thin film of
La$_{0.67}$Ca$_{0.33}$MnO$_3$ fabricated by a simple metalorganic
 decomposition technique $^{24}$, but it contrasts with Asamitsu
and Tokura's result$^{25}$ and that of Cao $et ~al.^{26}$ in which
the carriers are holes at  very low temperatures below $T_c$.
Furthermore, it was reported that the Mn$_{3d}$ band with
electrons increases near the Fermi energy and that the O$_{2p}$
peak with holes decreases as the photon energy increases in  the
photoemission  experiment; this supports the negative
$R_H$.$^{27}$ Therefore, for LCMO2/12H and LCMO9/15H films the
measurement of the negative $R_H$ indicates that the value of the
LTCMR rather than the HTCMR is considerably lower. Here, in
measuring the $R_H$, the resistance measured along the applied
current direction was negative, which indicates that the $R_H$ is
caused by the anomalous Hall coefficient which is generally
larger than the ordinary Hall coefficient for ferromagnetic
materials.

We now discuss comments on Cao $et~al.$'s Hall-measurement data
$^{26}$ on which Ziese$^{28}$ commented. Cao $et~al^{29}$
responded to Ziese's comment. Effects of two phases in Cao's
$R_H$ data were not separated completely because the $R_H$ was
determined by Hall resistance in a $\rho_{xy}$ (or diagonal)
direction. In this case, when the sample is annealed at
900$^{\circ}$C for at least 10 hours, the $R_H$ can become
negative at low temperatures far below $T_c$. Cao's suggestion,
which includes two types of conduction mechanisms in LCMO, can be
explained by introducing a two-phase concept, as suggested in
this paper. However, Ziese did not mention a two-phase concept in
his comment. The analysis of data in which effects of two phases
are merged is very complicated as Ziese mentioned. For our
analysis, we selected data obtained from 30 samples.

For the LCMO3/1H film, the LTCMR, $x$-HTCMR, positive MR, and
$R_H$ are shown in Fig. 3.
 Magnetoresistances, $\rho_{xx}$ and $\rho_{yy}$, were measured with 50  ${\mu}$A from 5 to  400 K by the van
 der Pauw method. The LTCMR in $\rho_{yy}$ measured along the $y$-axis with a peak at 90 K and the
$x$-HTCMR with a peak
 at 210 K are well  separated, as shown in  Figs. 3 (a) and (b).  The LTCMR in $\rho_{yy}$ is much more
 sensitive to the magnetic field than the HTCMR. Evidently the $y$-HTCMR was overshadowed
  by the LTCMR in Fig. 3 (b). With a 500${\mu}$A current flow through the  film, a
screened $y$-HTCMR approximately three times the size of the
$x$-HTCMR was observed near  250 K in Fig. 4 (b). This  indicates
that the film  has two  phases and  that the  magnetic moment
 is anisotropic.  Peak positions of the HTCMR in Fig.  4 shifted from 210 K to 250 K, though the positions did not shift.
Identifying the cause of the shift remains an open problem in
this paper. The LTCMR is reduced with an increasing annealing time
but
 does not disappear because it might have arisen from a  non-perfect crystalline order due  to a
dislocation, a defect, an oxygen
 deficiency or compositional  inhomogeneity.$^{12}$  For example,  Jaime et  al.$^{30}$
measured the HTCMR
 without the LTCMR but also observed a  thermopower peak near 150 K corresponding to
an LTCMR for a MN6/48H film deposited with  a
La$_{0.67}$Ca$_{0.33}$MnO$_3$ target by
 laser ablation on MgO and annealed in oxygen for 48 hours, as shown in
Fig. 3 in their
 paper$^{30}$. Moreover, by the four-probe method with line electrodes for LCMO3/1H, the
HTCMR, which depends
 on the ferromagnetic moment,  was  not found but
instead the LTCMR was observed. This
 indicates that  the  HTCMR was screened  by the LTCMR because  the HTCMR is  much
smaller than the LTCMR as shown in Figs. 3  (a) and (b).

Figure 3 (c) shows the $R_H$ obtained by the van der Pauw method
in the diagonal direction of the film. The shape of the $R_H$ is
similar to that  of the LTCMR in Fig. 3 (b). No effect of the
HTCMR  is seen because the small Hall voltage is screened by the
large Hall voltage induced by the LTCMR. The positive value of
the $R_H$ indicates that carriers are holes and corresponds to the
ordinary Hall coefficient because the semiconducting phase is not
ferromagnetic. The carrier density of approximately
1.2${\times}$10$^{15}$ cm$^{-3}$ at 90 K increased exponentially
up  to 2.2${\times}$10$^{17}$ cm$^{-3}$ at 400  K. These densities
were calculated by assuming that the effective mass  is equal to
the bare electron mass. The activation energy of the resistivity
is 61 $meV$ which is calculated from the $R_H$ data from 90 K to
400 K in Fig. 3 (c). This energy, due to excitation by thermal
phonons, can not be observed by optical methods. Thus, the
activation energy is  not a broad mid-infrared peak  observed by
optical methods$^{20,21}$  and
 a  minority spin  gap (${\approx}0.6 eV$)  observed by  spin-resolved  photoemission$^{14}$,
 but instead corresponds to a pseudogap. Calculated  from the
resistivity and  the $R_H$ measured  at  5 Tesla,  the mobility
is below 4  $cm^2/Vs$  and decreases overall with an increasing
temperature except for a broad convex curvature near 160 K, as
shown in the inset of Fig. 3 (c). This indicates that the
conduction is polaronic below and above 90  K. The polaronic
state observed by  Zhou et al.$^{11}$  corresponds to  this LTCMR
with hole carriers.

Figure 3 (d) shows remnant resistivities measured along the x-axis
as a function of field at $\pm$field at   5 K   and 90   K. The
resistivities are of  the  order  of 10$^{-3}~m{\Omega}cm$ (or
10$^{-8} ~{\Omega}m$)  below 100  K, as indicated by  A  in  Fig.
3  (a).  They are  nearly constant below  1  Tesla and   increase
with an increasing field above 1 Tesla (positive MR), except for
peaks near 400 Oe for resistivities at 90 K. This positive MR is
the same as that observed for the LCMO8/10H as shown at A in Fig.
1.

Figure 5 shows temperature dependences of resistances measured
along the x-axis and y-axis by the four-probe method with line
electrodes without magnetic field for an LSMO/MgO film deposited
with a La$_{0.67}$Sr$_{0.33}$MnO$_3$ target by laser ablation.
The current between electrodes was 50${\mu}$A. The film was not
annealed at a high temperature. The dimension of the film was
5${\times}$5 mm$^2$. The resistance $\rho_{xx}$ was measured
along the x-axis by the four-probe method with the y-axis-line
electrodes deposited with gold, as shown in the inset
 of Fig. 5.  After measuring $\rho_{xx}$, the gold electrodes on the
 film were removed with an etchant KI (Potassium Iodine) solution.
The surface of the film might have been slightly damaged, but the
damage was not obvious. After removing the y-axis-line electrodes
completely, the x-axis-line
 electrodes were deposited to measure $\rho_{yy}$. The resistance $\rho_{yy}$ was measured
along the y-axis by the same method as $\rho_{xx}$.  The
resistances have peaks A and C near 70 K corresponding to the
LTCMR and peaks B and D near 150 K corresponding to the HTCMR.
The peaks of the HTCMR were screened below 100 K by the peaks of
the LTCMR. Thus, low
 temperature information of the HTCMR is not revealed here.
This is different from the resistances shown in Fig. 1, Figs. 3
(a) and (b), and Fig. 4.

The two-phase concept was introduced to justify the van der Pauw
method. This concept has generally been accepted for oxide
materials. Some examples have been suggested as follows. Scanning
tunneling microscopy observed microscopic electronic
inhomogeneity on the surprisingly short scale of approximately
30$\AA$ for high-$T_c$ superconductors.$^{31-33}$ Metal-insulator
instability$^{34}$ near an optimal doping was theoretically
found. Heat capacity anomalies for strongly correlated metals
were explained by means of measurement.$^{35}$ Many papers
reported the phase separation in CMR materials,$^{13}$ which
indicate that the existence of two phases is intrinsic.
Therefore, the four-probe method with line electrodes, measuring
the resistance where the two-phase effects of metallic and
semiconducting properties are merged, is less
 effective for magnetic materials than the van der Pauw method  which is known to be
 applicable to homogeneous materials.

\section{CONCLUSION}
When magnetic moments are anisotropic and current flows  along a
direction with a large spin polarization, the resistivity of  the
ferromagnetic phase reaches a minimum. However, the minimum
resistivity is screened by the LTCMR when the LTCMR and the HTCMR
are not separated.

The semiconducting phase of the LTCMR is regarded as an impurity
phase imbedded in the ferromagnetic phase with the HTCMR. Thus,
this LTCMR phase is attributed not to the semiconducting
electronic structure predicted in a half-metallic electronic
structure$^{14}$ but to the remnant phase due to the
metal-insulator instability near an optimal doping$^{33}$.

The low resistivity, the  positive CMR, and the non-metallic
temperature   dependence of both the resistivity and the Hall
coefficient below $T_c$ indicate that ferromagnetic conduction is
not governed by free carriers, but instead by mobile polarons
condensed by an attractive potential energy due to a strong
electron-phonon interaction known as the Jahn-Teller effect$^7$.

Other possibilities for condensation of the ferromagnetic phase
have been presented as follows. A structure-phase transition with
a magnetic transition, caused by a large breathing-mode
distortion, was observed.$^8$ Electronic specific heat anomalies
were observed for a ceramic La$_{0.67}$Ca$_{0.33}$MnO$_3$ by
Ramirez et al al.$^{36}$ and La$_{0.8}$Ca$_{0.2}$MnO$_3$ by
Tanaka and Mitsuhasi$^{37}$, who suggested the anomaly was
evidence of a second-order transition. The density of states
(DOS) of the ferromagnetic phase, as the conductance of a tunnel
junction, was measured for La$_{0.67}$Sr$_{0.33}$MnO$_3$ at 4.2 K
by Yu Lu et al.$^{38}$ (see FIG. 3 in Yu Lu's paper). The DOS has
a parabolic shape within $\pm$0.2V and $V$ shape within
$\pm$0.05V regarded as the effect of LTCMR, indicating that when
the effect of LTCMR is removed, quasiparticles (or carriers) exist
not near the Fermi surface but below it. The same experimental
result was also observed by high resolution photoemission
spectroscopy by Park et al.$^{39}$ (see FIG. 3(c) in Park's
paper). The parabolic DOS is similar to the DOS of a
ferromagnet$^{40,41}$, a half metal$^{40,41}$, and a p-wave
superconductor as the ABM state$^{42}$.

Furthermore, we suggest that both the LTCMR and the van
 der Pauw method can be applied to a magnetoresistance sensor.\\

\begin{center}
\noindent{\bf ACKNOWLEDGEMENTS}
\end{center}
We acknowledge Ms. Juli Scherer as an editorial supervision of the
ETRI journal for proofreading.

\begin{figure}
\vspace{4.0cm}
\centerline{\epsfysize=8.0cm\epsfxsize=8.0cm\epsfbox{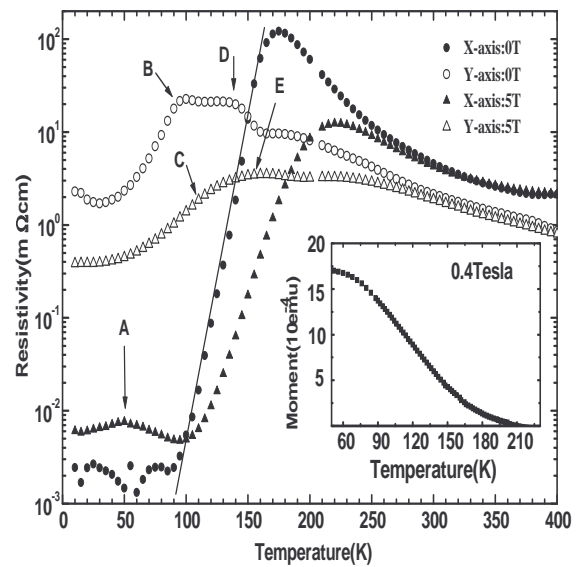}}
\vspace{-2.3cm} \caption{Temperature  and magnetic  field
dependencies  of magnetoresistivities  for the LCMO8/10H film. The
inset    shows the    temperature   dependence   of    the  z-axis
magnetic moment.}
\end{figure}

\begin{figure}
\vspace{-2.0cm}
\centerline{\epsfysize=9.0cm\epsfxsize=7.5cm\epsfbox{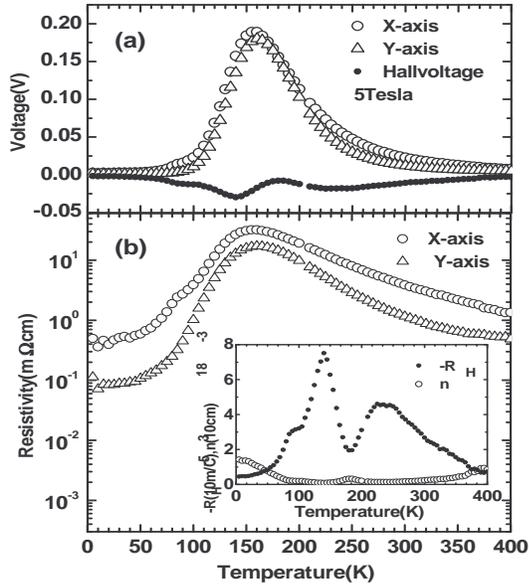}}
\vspace{0.0cm} \caption{(a) Temperature dependencies of
resistance and Hall voltages for the LCMO2/4H film. (b)
Temperature and magnetic field dependencies of
magnetoresistivities. The inset in Fig. (b) shows the Hall
coefficient and the carrier density.}
\end{figure}

\begin{figure}
\vspace{3.4cm}
\centerline{\epsfysize=8.5cm\epsfxsize=9.0cm\epsfbox{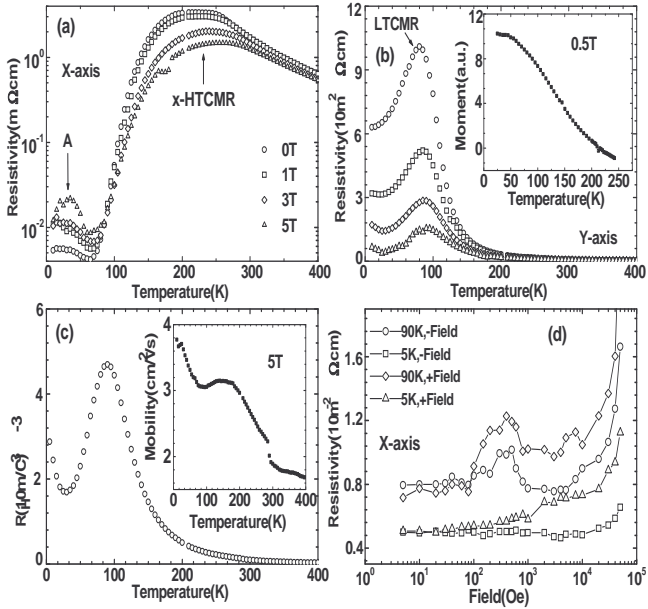}}
\vspace{-3.0cm} \caption{(a) and (b) Temperature  and  magnetic
field dependencies of magnetoresistivities  for  the LCMO3/1H
film. The inset in Fig. (b)   shows   the   temperature dependence
of the magnetic moment. (c) Temperature dependencies of the Hall
coefficient. The inset in Fig. (c) show  the mobility. (d)
Magnetic field dependencies  of magnetoresistivities measured at
+- fields at 5 K and 90 K for the LCMO3/1H film.}
\end{figure}

\begin{figure}
\vspace{0.1cm}
\centerline{\epsfxsize=7.0cm\epsfbox{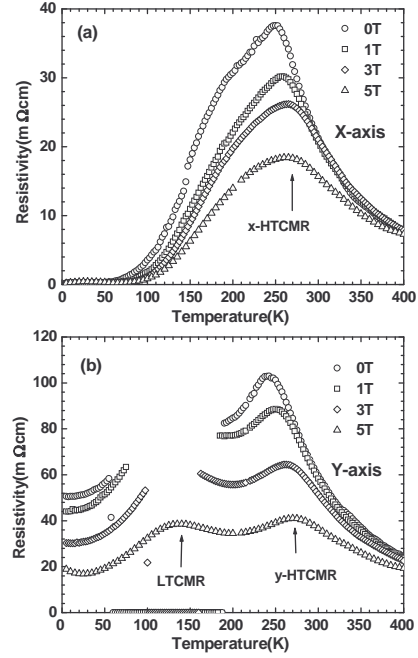}}
\vspace{0.1cm} \caption{Temperature and magnetic field
dependencies of magnetoresistivities measured with 500 ${\mu}$A
current for  the LCMO3/1H film. (a) Peaks corresponding  to the
x-HTCMR. (b) Peaks near 250 K and 100 K in Fig. (b) correspond to
the y-HTCMR and  LTCMR, respectively. Some data at fields of 0, 1
and 3 Tesla near
 100 K in Fig.(b) could not be measured  because of a measurement limit  of
 a voltmeter at 500 ${\mu}$A.}
\end{figure}

\begin{figure}
\vspace{2.5cm}
\centerline{\epsfxsize=8.0cm\epsfbox{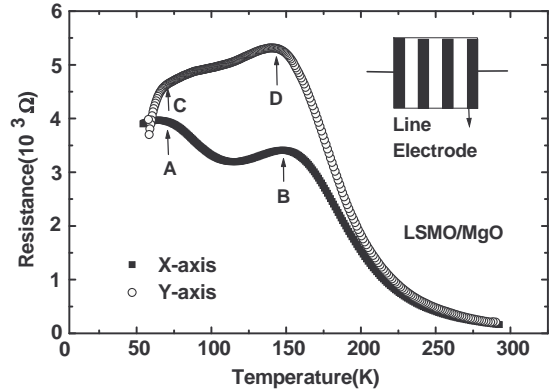}}
\vspace{-1.5cm} \caption{Temperature dependences of resistances
measured by the four-probe method with line
 electrodes at 0 Tesla with flowing 50 ${\mu}$A  for  the LSMO/MgO film. Peaks B and D
 correspond to the x-HTCMR and y-HTCMR, respectively. Peaks A and C correspond to the LTCMR.}
\end{figure}


\end{document}